 \documentclass[final,5p,times,authoryear]{elsarticle}

\usepackage{amssymb}
\usepackage{lipsum}
\usepackage{amsmath}

\journal{Physics Open}

\begin{document}

\begin{frontmatter}



\title{Machine Learning For Beamline Steering}


\author{Isaac Kante}
\ead{isaac.kante@my.liu.edu}
\affiliation{organization={Long Island University},
            addressline={1 University Plaza}, 
            city={Brooklyn},
            postcode={11201}, 
            state={New York},
            country={USA}}

\begin{abstract}
Beam steering is the process involving the calibration of the angle and the position at which a particle accelerator's electron beam is incident upon the x‐ray target with respect to the rotation axis of the collimator. Beam Steering is an essential task for light sources. In the case under study, the LINAC To Undulator (LTU) section of the beamline is difficult to aim. Each use of the accelerator requires re-calibration of the magnets in this section. This involves a substantial amount of time and effort from human operators, while reducing scientific throughput of the light source. We investigate the use of deep neural networks to assist in this task. The deep learning models are trained on archival data and then validated on simulation data. The performance of the deep learning model is contrasted against that of trained human operators.
\end{abstract}




\end{frontmatter}




\section{Introduction}
\label{introduction}

Beam steering \citep{metcalfe2007physics, gao2018quantification} is a process involving the calibration of the angle and the position at which a particle accelerator's electron beam is incident upon the x‐ray target with respect to the rotation axis of the collimator. The shape of the profile of the dose is extremely dependent upon accurate beam steering. This is essential to ensure correct delivery of the radiotherapy treatment plan. Beamline steering is essential in particle accelerator systems. These systems rely on beamline steering because it ensures that the beam stays on it's intended target and trajectories, which is critical for experimental and operation results. Beamline steering is important for several reasons like minimizing beam loss, maintaining beam stability, reducing interactions and accelerator components, error correction, and safety. It has a significant impact on the quality of scientific research, the safety of accelerator operations, and the success of various applications across multiple fields. Furthermore, beam steering is critical in mitigating errors and imperfections in the accelerator's magnetic and radio frequency systems, allowing for optimal performance and the successful execution of a variety of experiments and applications in fields such as particle physics, medical therapy, and materials science.

There are various methods used for beamline steering. I discuss the primary ones:\\
\textit{Steering magnets} - This method includes dipole and quadropole magnets that are used to adjust the trajectory of the beam. These magnets generate magnetic fields that deflect or focus the beams as needed. Precise control of the magnetic field strength and alignment allows for accurate steering \citep{steeringmagnets}.\\
\textit{Beam Position Monitors (BPMs)} - This methods includes placing bpms along the beamline to detect and measure the position of the beam. The data from the bpms are used to make adjustments (\cite{Bpm}).\\
\textit{Orbit Correction Systems} - Orbit correction systems consist of a combination of steering magnets and position monitors. They continuously monitor the beam's position and adjust the magnets' settings to keep the beam on its desired trajectory, making real-time corrections (\cite{orbital}).\\
\textit{Reducing Interactions with Accelerator Components} - When beams stray from their planned routes, they might contact with accelerator components, resulting in undesirable consequences such as heating, radiation, or damage. Precision steering reduces these interactions, extending the life of accelerator components.

Machine learning offers a good alternative for beamline steering due to it's ability to handle complex and dynamic real-time data. This allows for precise adjustments to maintain beam stability and optimize trajectory. Machine learning offers processing vast of information such as bpms, and learn patterns that humans might overlook. Machine learning algorithms can assess this data in real time, discover deviations from the expected trajectory, and generate the ideal settings for steering magnets and other beamline components on their own. They can adapt to changing conditions and respond to disturbances more quickly, enhancing beam quality and lowering the chance of beam loss.

Some applications of Machine Learning in Beamlines include anomaly detection, system modeling, virtual diagnostics, data analysis, and beamline tuning.

\textit{Anomaly detection and machine protection} - Machine learning models may identify abnormalities or unexpected behaviors in accelerator systems in real time, such as beam instabilities or equipment faults \citep{fol2017evaluation}. When abnormalities are discovered, these systems can initiate safety measures or shutdown processes to safeguard equipment and workers.

\textit{System Modeling} - Machine learning is used to create models that replicate accelerator system behavior, assisting in the prediction and understanding of how different parameters and components impact beam trajectories and performance (\cite{gupta2021improving, mishra2021uncertainty}). These models help in the design and optimization of systems.

\textit{Virtual Instrumentation and Virtual Diagnostics}- Machine learning is used in virtual instrumentation and diagnostics to construct virtual representations of accelerator components and systems. These simulations can give insights on beamline behavior, allowing for the testing of different circumstances without the need for real tests, thereby assisting in system design and diagnostics \citep{pang2015advances}.

\textit{Tuning, Control, Rapid Switching Between Operating Conditions}- To ensure optimal beam steering, machine learning techniques are used to automatically tune and manage accelerator settings \citep{scheinker2015adaptive}. They can adjust to changing conditions in real time and allow for quick switching between operational modes to fulfill experimental or operational requirements.

\textit{Advanced Data Analysis} - To extract important insights from the massive amounts of data generated by accelerator systems, machine learning and sophisticated data analysis techniques are applied. They can spot patterns, correlations, and trends in data to enhance beamline guidance, solve problems, and boost overall performance.

These applications highlight how machine learning could improve the efficiency, precision, and automation of beamline steering and accelerator operations, resulting in better experimental findings, less downtime, and increased safety. The area is evolving as a result of advances in machine learning techniques and the incorporation of artificial intelligence into accelerator facilities (\cite{edelen2018opportunities}).

\section{Mathematical, Dataset And Numerical Details}

\subsection{Deep Learning Based Modeling}

Deep learning is a form of machine learning that uses neural networks with numerous layers to uncover complicated patterns from difficult data. These deep neural networks, also known as deep learning models, have shown to be extremely useful in a variety of disciplines (\cite{10056957}). Deep learning excels at picture and audio recognition, natural language processing, and other tasks (\cite{Deeplearning}). It is based on an architecture inspired by the biological neurons found in the brains of humans, that employs layers of linked neurons to gradually improve and abstract data representations. This layer structure enables deep learning models to automatically learn and adapt to data attributes, resulting in cutting-edge outcomes in tasks such as picture classification, object identification, and language translation. Figure \ref{model} is the deep learning model used for this project with seven fully connected layers, each followed by a Rectified Linear Unit (ReLU) activation function. It takes a certain size input, runs it through numerous hidden layers, and returns predictions. The model can capture detailed patterns and correlations in the data due to this architecture. The use of ReLU activation promotes nonlinearity, which improves the model's ability to learn complex mappings. Because of its modular nature the deep learning model is adaptable and flexible, making it suited for a variety of applications such as regression or classification. In this study we use the PyTorch library (\cite{NEURIPS2019_9015}) to create, test and tune the deep learning models.

\begin{figure*}
    \centering
    \includegraphics[width=\textwidth, trim={17cm 5cm 8cm 5cm},clip]{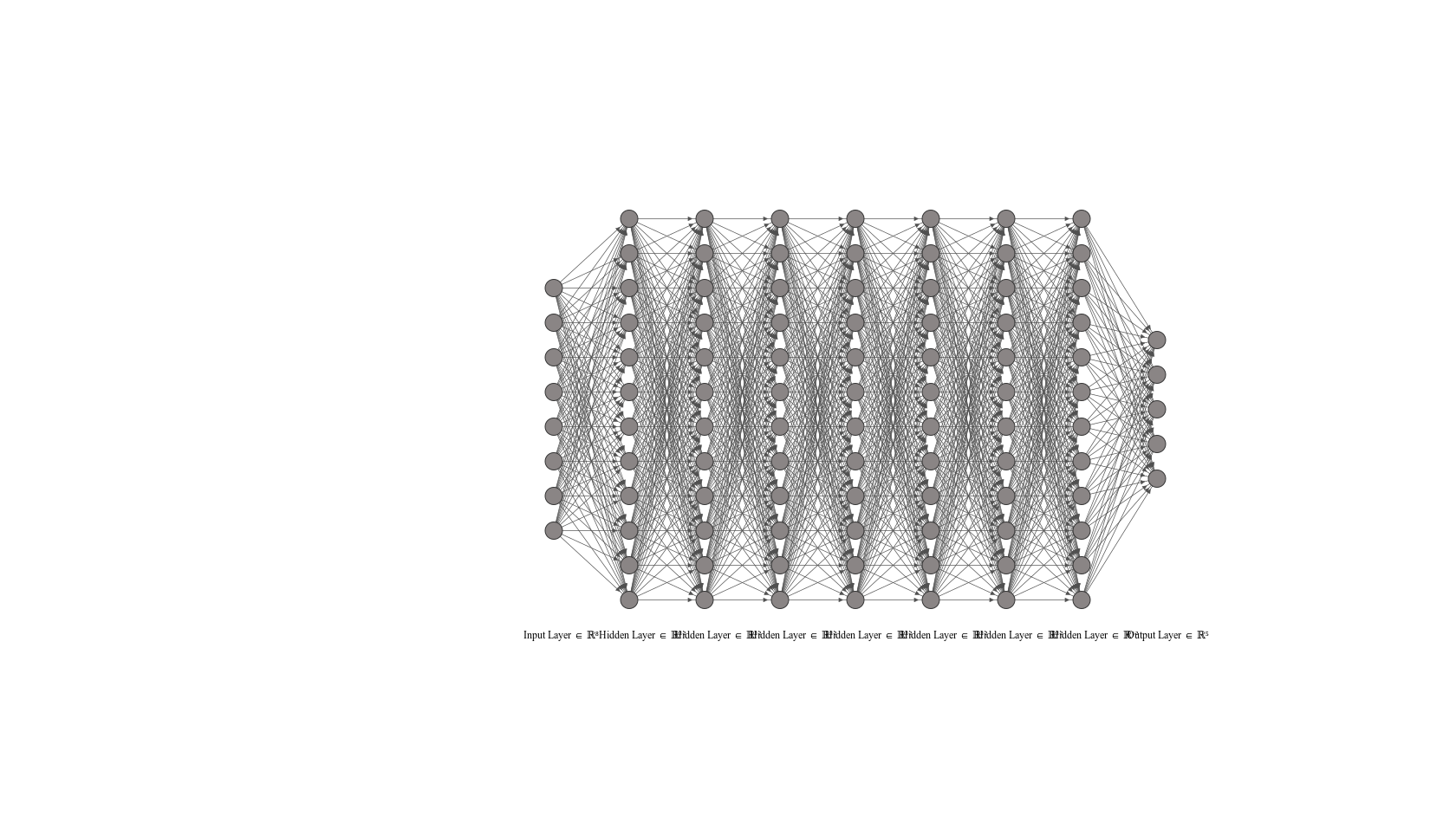}
    \caption{Neural Network Architecture used in this study. The Input is 73 dimensional, each of the hidden layers have 512 units, the output is 50 dimensional.}
    \label{model}
\end{figure*}

\subsection{Dataset Details}

\begin{figure*}
    \centering
    \includegraphics[width=\textwidth, trim={0cm 5cm 0cm 5cm},clip]{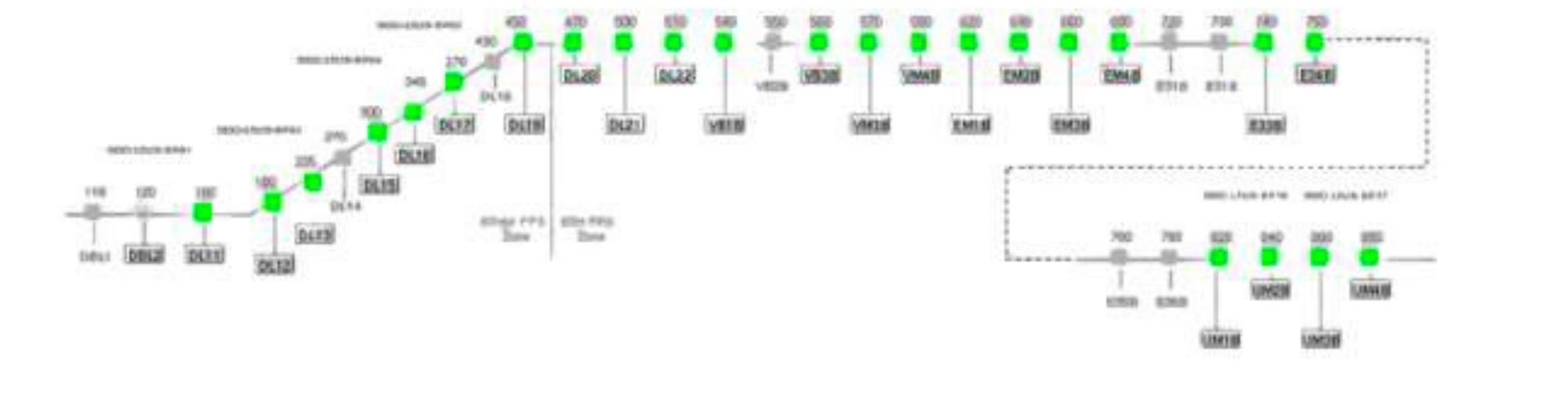}
    \caption{Schematic of the Beamline, highlighting the LTU (LINAC To Undulator) section. This section is difficult to correct for the operators. Each use of the accelerator requires recalibrating the magnets in this section.}
    \label{model}
\end{figure*}

The Linac Coherent Light Source (LCLS) is a x ray Free Electron Laser (FEL) based on a long, permanent magnet undulator constituted of numerous segments that provide space for electron beam position monitors and quadrupole focusing magnets \citep{emma1992beam}. This is schematically shown in Figure 2. The 15 GeV electron trajectory within the undulator must be completely straight to $5\mu m$ over a gain length of 11 m so that the photon beam overlaps efficiently with the electron beam. To train the model, we use archival data with the required inputs and outputs for beam steering. The data being used for this project are different Process Variables (PVs) arrays including XCOR, YCOR, QUAD, BPMX, and BPMY.

The input features are X-Correcter (XCOR), Y-Corrector (YCOR), and Quadropole magnet settings (QUAD).\\
XCOR and YCOR are horizontal and vertical beam correctors which are utilized for shifting and adjusting the particle beam inside the beamline.\\
QUAD is the quadropole magnet that consist of groups of four magnets laid out, which are used for focusing the particle beam. Bar magnets have 2 poles and are di-polar. Quadrropole magnets have 4 poles. They can be used to focus and transport charge particle beams, like x-rays, through long beamlines, in linear colliders and circular machines like synchrotrons. \\
BPMX and BPMY are refered to as the "Beam Position Monitors" in the horizontal and vertical directions. Beam Position Monitors (BPM) are the non-destructive diagnostics used most frequently at nearly all linacs, cyclotrons, and synchrotrons. BPMs deliver the centre of mass of the beam and act as a monitor for the longitudinal bunch shape. BPMs are devices that measure the location of the charged particle beam as it goes through the beamline in particle accelerators. They give real-time input on the location of the beam, allowing for fine control and modifications, including trajectory correction. Each of these fields represents a type of PV and contains an object with value arrays.

\subsection{Hyperparameter Selection}
In this section I want to highlight the selection of the hyperparameters and the approach of metric being used. The deep learning model architecture being used in this project was a feedforward neural network. This consisted of 7 linear layers with a loss function of mean squared error (MSE) to calculate the loss for this regression task. The layers take in an input, hidden layers and an output. A forward pass method than specifies the data through the network with a Rectified Linear Unit (ReLU) activation in sequence \cite{Relu}. The training loop hyperparameters used for the model was a learning rate set to 0.001, which is used for controlling the step size taken during optimization \cite{learningrate}. The optimization algorithm used was the Adam which is short for Adaptive Moment Estimation \cite{Adam}. The training loop was set to an epoch of 300. This parameter is the amount of times the model would see the data during training \cite{sinha2010epoch}. The patience parameter is set to 5 for early stopping if the validation loss does not improve for 5 consecutive epochs. This parameters helps with overfitting and saving training time \cite{earlystopping}.

\subsection{Simulation based data generation}

The program used for generating simulation data is Bmad the software toolkit for charged particles and x-ray simulations. According to \cite{Bmad} "Bmad can be used to study both single and multi–particle beam dynamics as well as X-rays". Bmad is not a program itself and is used by programs. The program used was Taothat was developed around bmad for this reason. "Tao is a general purpose simulation program, based upon Bmad. Tao can be used to view lattices, do Twiss and orbit calculations, nonlinear optimization on lattices, etc \cite{Tao}.

Ensuring the simulation data corresponds to the same exact setting were factored by certain steps. Keeping the data structure the same such as the file and inputs that were used in model training, be the same when generating the data. Keeping the same number of features as the training data and making sure the targets are the corresponding targets as the training data. Applying the same pre-processing steps as the training ensures the same settings as the model. Evaluation metrics MSE and MAE stay the same for both the simulation and real data. The amount of simulation data that was generated for testing were 6 data sets. 

\subsection{Archival data}

The archival data was derived from EPICS (Experimental Physics and Industrial Control System) archiver. Specific timestamps from where we want the data are included as well as the format we want the data in. Data retrieval involved calling the method lcls archiver restore and supplies the parameters PVLIST and epics timestamp. The function interacts with the EPICS archiver to get historical data for the Process Variables (PVs) mentioned in PVLIST at the supplied timestamp.

\section{Hyperparameter Tuning}

Tuning hyperparameters is an important part of developing machine learning models since it has a direct influence on model performance and generalization. Hyperparameters are settings that cannot be learnt from data and must be specified beforehand. Learning rates, batch sizes, network designs, regularization settings, and other factors are among them. Properly calibrated hyperparameters can result in a model that not only fits the training data well but also generalizes well to previously unknown data. According to \citep{hyperparameter}  “Bayesian optimization found a better instantiation of nine convolutional network hyperparameters than a domain expert, thereby achieving the lowest error reported on the CIFAR-10 benchmark at the time”. The goal of hyperparameter tuning is to optimize the hyperparameters in a way that improves the performance of the machine learning or deep learning model. 

\subsection{Automating Hyperparameter Tuning} 

Because manual exploration of hyperparameter settings is time-consuming as well as inefficient, automating hyperparameter tweaking is essential. Grid search, random search, and Bayesian optimization are examples of automated algorithms that systematically examine the hyperparameter space to discover the optimal combination for the given job. These approaches involve assessing the model's performance on validation data for various hyperparameter settings, allowing optimal values to be selected without substantial user involvement. Automated hyperparameter tweaking can save time and increase model performance in complicated settings with many hyperparameters. It also makes repeatability and the capacity to adjust models to changing data easier. One of the automated hyperparameter frameworks I have used for this project is Optuna. Optuna is a framework for automated hyperparameter optimization that simplifies the process of choosing optimal hyperparameter settings for machine learning models. It automatically explores through a set of hyperparameter values, determined by optimization algorithms, in order to optimize a certain objective function \citep{optuna}. Optuna has 3 design criteria that are made to outperform many major black-box optimization frameworks while being easy to use and easy to setup invarious environments \citep{optuna}. Define-by-Run Programming: This means that you can dynamically create and change the search space for hyperparameter optimization as your program runs. It allows you to specify, modify, or extend the hyperparameters you want to optimize without requiring a fixed, predefined configuration. Efficient Sampling and Pruning: The framework offers efficient methods for trying out different hyperparameter values (sampling) and allows you to customize how these values are explored. It also supports a technique called pruning to stop unpromising trials early, making the optimization process more efficient. Versatile and Easy to Use: The framework is designed to be flexible and adaptable for various tasks. It can be easily set up, from simple interactive experiments to complex distributed computations, making it a versatile tool for optimizing machine learning models. In Figure\ref{optuna} I demonstrate the use of Optuna in optimizing a neural network model. The figure illustrates the model architecture, showcasing its dynamic layers and the Optuna-driven objective function for hyperparameter tuning. The optimization explores the search space for hidden dimensions, dropout rates, and learning rates. This approach efficiently refines the model's performance through iterative tuning, resulting in improved accuracy. The flexibility and adaptability of Optuna facilitate a systematic exploration of hyperparameter configurations, enhancing the overall effectiveness of the machine learning model.

\section{Model Prediction}

In this section I compare the the models predictions against the ground truth of the data. This evaluation is essential to assessing the model's accuracy and dependability in real-world circumstances. We acquire vital information about its capabilities and limits by comparing its performance against established benchmarks. This study not only confirms the model's reliability, but also encourages prospective changes, resulting in a strong and trustworthy prediction architecture. By basing our analysis on actual data, we improve the application and relevance of our conclusions, thereby improving the overall credibility of our prediction model. 
\subsection{Archival and Simulation Data}
I use four datasets for evaluation: 1 archival and 3 simulation. The archival dataset is made up of real-world, historical data that captures genuine system activity. Meanwhile, the simulation dataset contains artificially generated data that simulates various scenarios. The strength of archive data is its authenticity, which reflects true system details. It may, however, be lacking in diversity. However, simulation data assures a wide range of scenarios but may not fully reflect real-world details. Balancing various datasets eliminates any biases and improves the model's flexibility to varied conditions, resulting in a complete assessment technique. Juggling the authenticity of archival data with the diversity of simulated scenarios for a comprehensive evaluation is one of the challenges. We use key metrics to evaluate the model's prediction accuracy, such as Mean Squared Error (MSE), Root Mean Squared Error (RMSE), and Mean Absolute Error (MAE). MSE calculates the average squared difference between expected and actual values, favoring precise forecasts while punishing bigger mistakes. In the original scale, RMSE, the square root of MSE, gives a more interpretable statistic. MAE calculates the average absolute difference, which provides a reliable measurement of prediction accuracy. These measures were chosen because of their ability to capture many elements of model performance while addressing the problem's sensitivity to both tiny and big mistakes. RMSE and MAE provide insight into prediction accuracy and help in determining the model's generalization across different targets, providing a thorough assessment technique for our regression problem.

\subsection{Visualization Prediction Assessment}
\begin{figure*}
	\centering 
	\includegraphics[width=\textwidth, angle=0]{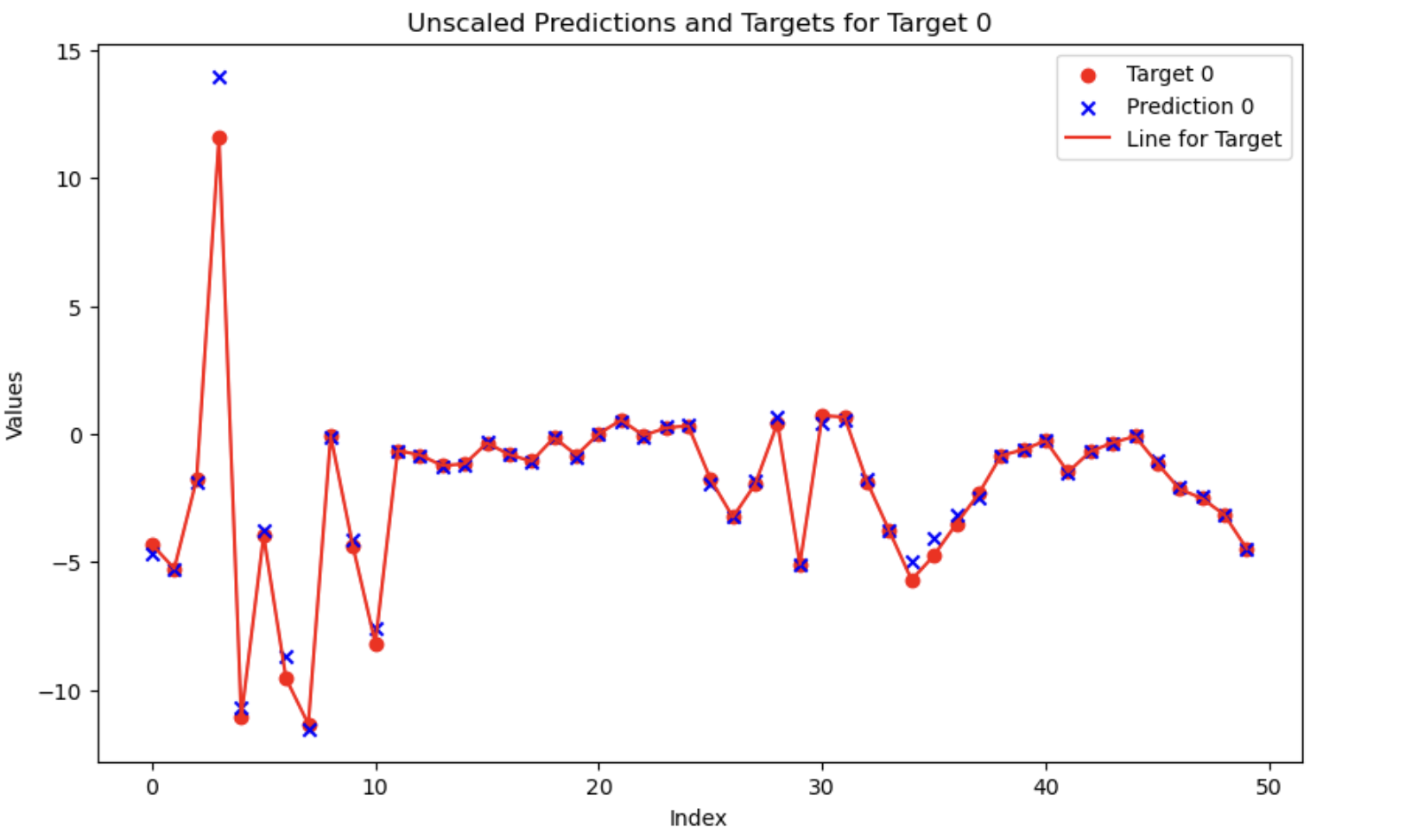}
        \centering
        \caption{Model prediction vs Truths on Archival Data } 
        \label{archival data}%
\end{figure*}

I focused our investigation on 4 targets, assessing its predictions across four datasets. These datasets included one from the archival data sources and three from simulations. I hoped to gain a more detailed understanding of how our model works on various data sources by isolating a certain goal. Predictions using archive data provide insights on real-world applicability, whilst simulated datasets allow examination of the model's adaptability to diverse settings. The comparison across datasets improves our understanding of the model's resilience and generalization. In Figure \ref{archival data} the model shows its predictive accuracy with the archival data. An abundunce of the model predictions vs the ground truth are aligned closely as shown in the plot. The model shows its potential for high efficiency prediction. 

\begin{figure*}
	\centering 
	\includegraphics[width=0.65\textwidth, angle=0]{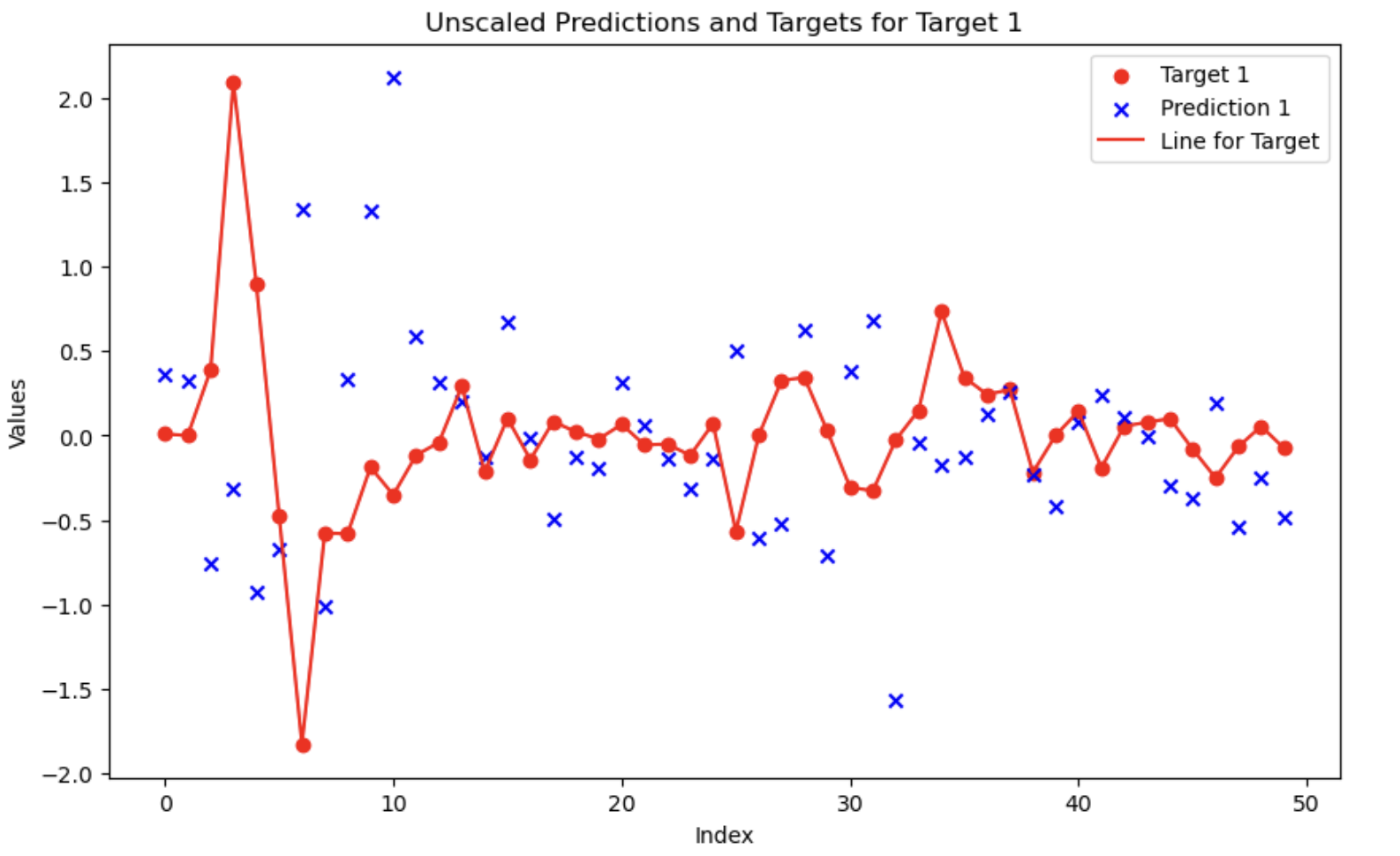}
        \centering
        \includegraphics[width=0.65\textwidth, angle=0]{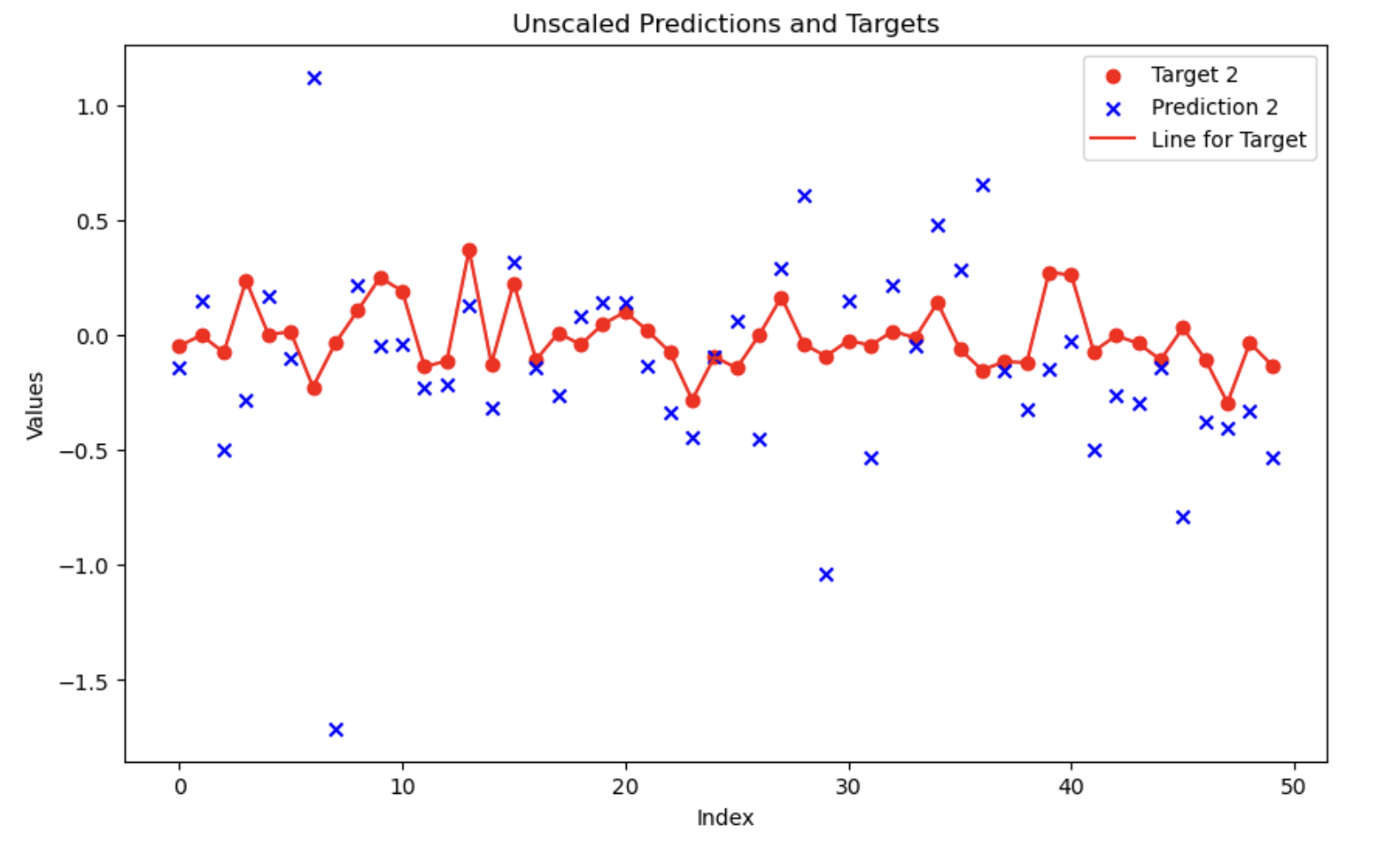}
        \centering
        \includegraphics[width=0.65\textwidth, angle=0]{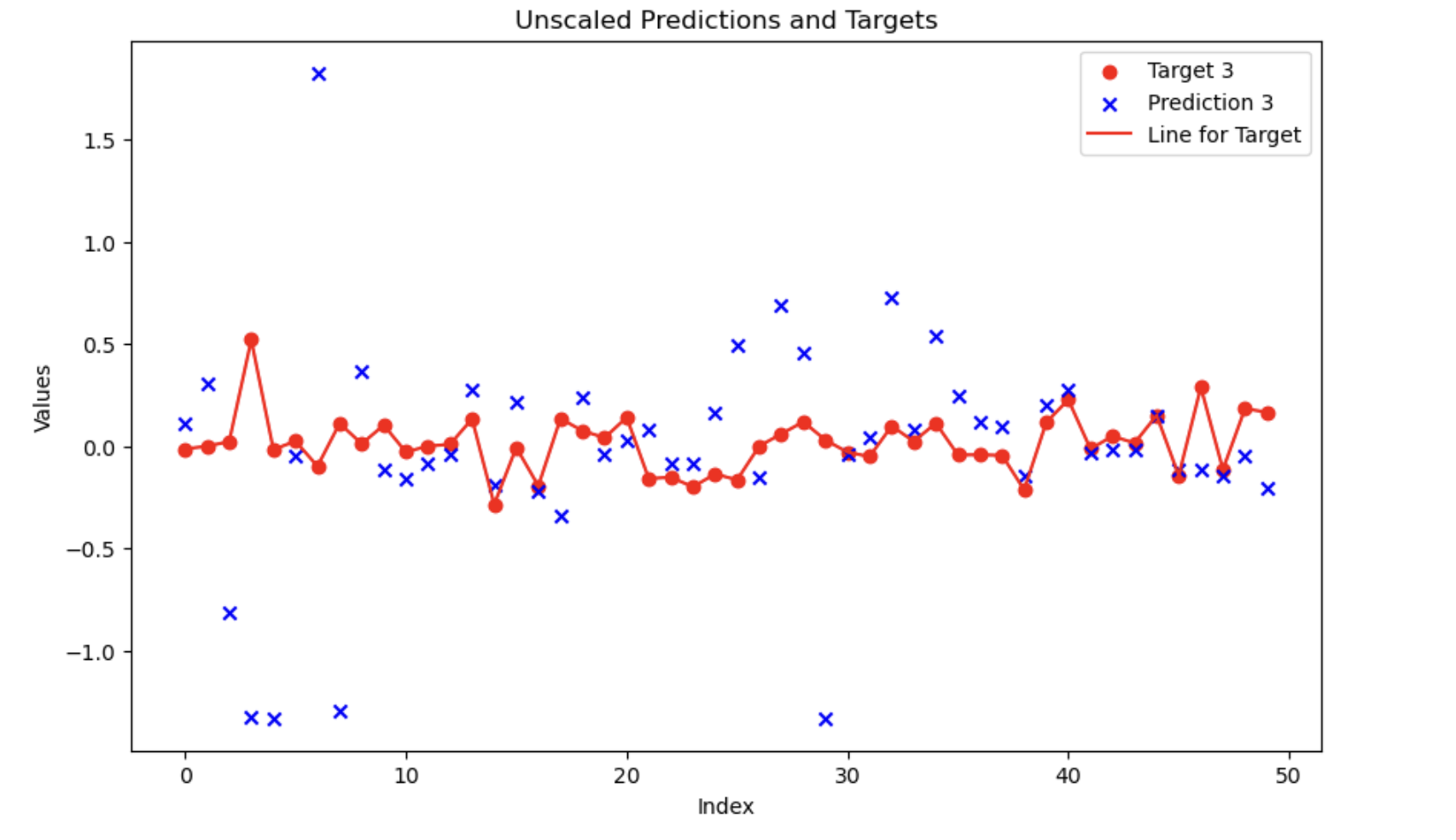}
        \caption{Model prediction vs Truths on Simulation Datasets} 
        \label{simlation data}%
\end{figure*}

The scatter plots in Figure \ref{simlation data} show a significant difference between model predictions and ground truth for simulated data. The scattered distribution of points indicates difficulties in the model's adaptability to the complexities of simulated events. Disagreements may arise as a result of the small properties presented in the simulation environment, influencing the model's generalization. This visualization serves as an important diagnostic tool, identifying areas where the model may need to be improved which would be beneficial. Addressing these disparities in simulated data is consistent with ongoing work to improve the model's resilience, assuring its efficacy across various datasets and confirming its trustworthiness in real-world applications.

\section{Performance}

In this section, we examine our model's performance across multiple datasets. First, we evaluate its performance on archival data, gaining insights into its accuracy and generalization. Following that, the model's performance on simulated datasets is examined, revealing information on its adaptability to various settings. The verification procedure assures that the model is reliable and consistent. I hope to give a full review of our model's strengths and limits in collecting patterns across both real-world and simulated environments by thoroughly analyzing metrics and visualizations.

\subsection{Model performance on Archival data}
The evaluations provide important insights into the performance of our model on archival data. As shown in Table \ref{TABLE 1} the Mean Absolute Error (MAE) of 0.159 shows that our model's predictions differ from the actual values by 0.159 units on average. At 0.308, the Mean Squared Error (MSE) indicates the average squared difference between forecasts and true values, emphasizing the importance of greater mistakes. The Root Mean Squared Error (RMSE) of 0.555, which is obtained from the square root of the MSE, provides a measure of the normal amount of errors, providing a comprehensive picture of model accuracy.
The low values metrics for archival data suggest that our model performs decently on the dataset. The closeness of predicted and actual values indicates the model's potency in capturing patterns in archival data. The model's robustness is demonstrated to be inadequate in the field of new generalization to new examples. A careful review of these indicators ensures a thorough grasp of our model's capabilities on the archival dataset.
\subsection{Model performance on simulation data}

The notable difference in model performance between archival and simulation datasets raises concerns. While the model performs admirably on archive data, larger errors on Simulation Data 1, 2, and 3 in Figure \ref{TABLE 1} indicate difficulties in adjusting to new settings. Among the simulated datasets, Simulation Data 1 has the largest Mean Absolute Error (MAE), Mean Squared Error (MSE), and Root Mean Squared Error (RMSE). This imbalance highlights potential differences between the simulated and real-world data distributions. This discrepancy may be caused by factors such as noise, parameter mismatch, model architecture, model interpretability, and data quality. Simulation Data 2 provides a decent MAE but a much lower RMSE, showing that outliers may be influencing the total error. It requires deeper examination into the error distribution and the model's sensitivity to extreme values.
Despite having a lower MAE, the model on Simulation Data 3 has a larger RMSE than Archival Data. Differences in data features, such as feature distributions and patterns, between archival and simulated datasets present a substantial difficulty. Overfitting of the model during training on more controlled archival data may contribute to its inability to generalize to the complexities presented by the simulated datasets.

\subsection{Model verification}

Model verification is an important step in ensuring the model's dependability and suitability for deployment. This procedure involves thorough testing and comparing model outputs to ground truth across multiple datasets. In our example, verification evaluates the model's performance consistency across archival and simulation data. Disagreements may signal the need for additional refining. We can learn about the model's resilience and accuracy by looking at measures like MAE, MSE, and RMSE. Successful verification boosts confidence in the model's ability to generalize, increasing trust in its predictions across a wide range of circumstances and datasets.

\begin{table}
    \caption{Model Performance Metrics on Archival data and on data from Bmad simulations.}
    \centering
    \begin{tabular}{l c c c} 
    \hline
    Data & MAE & MSE & RMSE\\ 
    \hline
    Archival Data     & 0.159 & 0.308 & 0.555 \\ 
    Simulation Data 1 & 0.606 & 0.816 & 0.903  \\
    Simulation Data 2 & 0.311 & 0.200 & 0.448  \\
    Simulation Data 3 & 0.340 & 0.324 & 0.569  \\
    \hline
    
    \end{tabular}
    \label{TABLE 1}
\end{table}

\section{Beamline Steering and Orbit Correction}
In the context of beamline steering and orbit correction for accelerators, the following principles are fundamental. The orbit response matrix, denoted as $\mathbf{R}$, plays a crucial role in characterizing the relationship between the desired changes to corrector magnets ($\boldsymbol{\theta}$) and the resulting difference between the target and measured orbits ($\Delta \mathbf{x}$). These concepts, central to the beam-based correction and optimization of accelerators, have been extensively explored by Huang in the book 'Beam-based correction and optimization for accelerators' (2020) \citep{beam}.

\subsection{SVD Approach}
The Singular Value Decomposition (SVD) of the orbit response matrix provides a key insight into the modes governing orbit changes. This method, as outlined by Huang, expresses $\mathbf{R}$ as $\mathbf{U} \boldsymbol{\Sigma} \mathbf{V}^T$, where $\mathbf{U}$ and $\mathbf{V}$ are orthogonal matrices, and $\boldsymbol{\Sigma}$ is a diagonal matrix of singular values. The effectiveness of different modes in inducing orbit changes is encapsulated in these singular values.

Throughout this discussion, we will delve into the application and significance of these principles, drawing on the comprehensive work presented by Huang \citep{beam}.

The SVD (Singular Value Decomposition) method for beamline steering is a powerful way to correct the path of particles in accelerators. In this method, the goal is to minimize the difference between the measured beam orbit and the target orbit in a least-square sense. The system involves multiple BPMs (Beam Position Monitors) and correctors. The key equation for orbit correction with SVD is:
\begin{equation}
    \theta = (R^T R)^{-1} R^T \Delta x
\end{equation}

Here, $\mathbf{R}$ is the orbit response matrix, $\boldsymbol{\theta}$ is the vector of desired changes to the corrector magnets, and $\Delta \mathbf{x}$ is the difference between the target orbit and the measured orbit.
The SVD of the orbit response matrix, as discussed in 'Beam-based correction and optimization for accelerators' \citep{beam}, is expressed as
\[
\mathbf{R} = \mathbf{U} \boldsymbol{\Sigma} \mathbf{V}^T,
\]
where $\mathbf{U}$ and $\mathbf{V}$ are orthogonal matrices, and $\boldsymbol{\Sigma}$ is a diagonal matrix of singular values. The singular values represent the effectiveness of different modes in making orbit changes. The SVD method provides a unique solution to the least-square problem even when the system is over-constrained or under-constrained.
The solution is further refined by considering the pseudo-inverse of $\mathbf{R}$, especially when the number of correctors is greater than the number of BPMs. The pseudo-inverse is calculated using the singular values, ensuring a stable solution. The SVD approach allows for weighting BPMs differently based on their importance in the orbit control system. This flexibility is valuable in situations where precise control is needed at specific locations along the beamline.
The method is demonstrated in practical scenarios, such as the correction of orbit errors in accelerators like SPEAR3. The results show that with a relatively small number of singular value modes, significant reduction in orbit distortion can be achieved. The approach is also extended to include weighting factors for BPMs, allowing for fine-tuning of orbit control based on specific requirements at different locations.

\subsection{SVD vs Deep Learning}
The Singular Value Decomposition (SVD) and Deep Learning represent distinct approaches with specific strengths and applications. SVD is well-suited for linear problems where the underlying physics or mathematics is well understood, making it effective for scenarios like beamline steering in accelerators. It provides clear insights into contributing modes and interpretable matrices. However, SVD's applicability diminishes in highly nonlinear systems.

Deep Learning excels in solving complex, nonlinear problems, offering accurate predictions where traditional models fall short. It operates as a "black box" lacking the interpretability of SVD. While SVD demands a grasp of physics and accurate data, Deep Learning thrives on large datasets, making it ideal for intricate relationships.

In terms of flexibility, SVD may struggle with dynamic systems, assuming a certain linearity, while Deep Learning adapts to diverse problems, suitable for evolving patterns. SVD relies on established mathematical procedures, eliminating the need for training, unlike Deep Learning, which requires extensive training and computational resources.

Application-wise, SVD finds use in physics, engineering, and linear systems, exemplified by beamline steering. Deep Learning dominates image recognition, natural language processing, and other areas with complex, nontraditional data. Both methods cater to specific niches, emphasizing the importance of selecting the right tool for the nature of the problem at hand.

\section{Conclusions And Future Work}
Beamline steering is an essential task for light sources. Currently beamline steering is carried out using trained operators. Faster, more reliable and automated beamline steering would be beneficial to light sources in increasing the capacity of experiments to be executed at the facility and the quality of the results. Machine Learning models are fitting for this application as we have a lot of data to learn this mapping. Deep learning based models with their accuracy and flexibility are the prime choice. Deep learning models sometimes give overconfident answers and are not immune to covariate shifts.
In the case under study we applied Deep Learning models for beamline steering. On the archival data where the models were trained the models gave very precise answers. But on simulation datasets their accuracy was off by a factor of 2 or more. This highlights the susceptibility of deep learning models to covariate shifts. In all the studies the deep learning models were almost an order of magnitude worse than the Singular Value Decomposition approach being used right now for beamline steering. 
In future work I recommend the use of Recurrent Neural Network based models. These models are used to model sequences where there is an order. On the beamline where different instruments are positioned at different locations one after the other, this sequence modeling should increase the quality of predictions by incorporating this inductive bias.

\bibliographystyle{elsarticle-harv}
\bibliography{mybibliography}






\end{document}